A study of the possibility of sprites in the atmospheres of other planets

Yoav Yair[1], Yukihiro Takahashi[2], Roy Yaniv[1,3], Ute Ebert[4,5] and Y. Goto[6]

1. Department of Life and Natural Sciences, The Open University of Israel, Israel
2. Department of Geophysics, Tohoku University, Sendai, Japan
3. Department of Geophysics and Planetary Sciences, Tel-Aviv University, Israel
4. Centrum Wiskunde & Informatica (CWI), Amsterdam, The Netherlands
5. Faculty of Physics, Eindhoven University of Technology, Eindhoven, The Netherlands
6. Department of Electrical Engineering, Tohoku Gakuin University, Tagajo, Japan

Contact Author: Yoav Yair

Department of Life and Natural Sciences

The Open University of Israel

108 Ravutski St., Ra'anana 43107

Israel

**Abstract**. Sprites are a spectacular type of transient luminous events (TLE) which occur above thunderstorms immediately after lightning. They have shapes of giant jellyfish, carrots or columns and last tens of milliseconds. In Earth's atmosphere, sprites mostly emit in red and blue wavelengths from excited $N_2$ and $N_2^+$ and span a vertical range between 50 and 90 km above the surface. The emission spectra, morphology and occurrence heights of sprites reflect the properties of the planetary atmosphere they inhabit and are related to the intensity of the initiating parent lightning.. This paper presents results of theoretical calculations of the expected occurrence heights of sprites above lightning discharges in the $CO_2$ atmosphere of Venus, the $N_2$ atmosphere of Titan and the $H_2$-He atmosphere of Jupiter. The expected emission features are presented and the potential of detecting sprites in planetary atmospheres by orbiting spacecraft is discussed.

1. **Background**

**1.1 Observations of Planetary Lightning Activity**

Apart from Earth, lightning has been discovered or inferred in several other planetary atmospheres in the solar system. We shall review these observations briefly, and would like to refer the interested reader to the extensive reviews by Desch et al. (2002), Aplin (2006) and Yair et al. (2008). On Venus, lightning activity had been deduced based on the VLF emission detected by the Soviet landers Venera 11 and 12 (Ksanfomality, 1980). However, the data from top-side observations by various spacecraft have not shown un-equivocal optical or electromagnetic signatures, especially after the fly-bys of the Galileo and Cassini spacecrafts (Gurnett et al., 1991; Gurnett et al., 2001). Krasnopolsky (2006) obtained high-resolution spectra of Venus in the NO band at 5.3 μm and found an NO content of 5.5±1.5 ppb below 60 km altitude. Such a concentration cannot be explained by cosmic-ray induced chemistry and the suggested mechanism is production by lightning. Lately, based on the Venus Express magnetometer data Russel et al. (2007) showed that lightning activity on Venus is of the same order of magnitude as on Earth, namely ~50 s$^{-1}$. There is still uncertainty concerning the global Venusian lightning rate in view of the fact that no plausible charge separation mechanism had been suggested to account for such a high flash rate (Levin et al., 1983). On Mars, there are presently no direct observations of electrical activity, but it is expected that the frequent small dust devils (z < 100 m) and larger storms (z ~ 10 km) would be electrified due to triboelectric charging processes and display some form of discharge (Melnik and Parrot, 1998; Farrell et al., 1999). While there were theoretical predictions for lightning activity in the convective methane clouds of Titan (Tokano et al., 2005), none were detected even after 17 close-range flybys by the Cassini spacecraft (Fischer et al., 2007). This may indicate that thunderstorms are exceptionally rare or even non-existent in that atmosphere, and attest to the weakness of charge build-up processes within the methane clouds. At the giant planets, the Voyager, Galileo, Cassini and New-Horizons missions found clear indications that lightning is prevalent on Jupiter (Magalheas and Borucki, 1991; Little et al., 1999; Baines et al., 2007) and also occurs on Saturn (Gurnett et al., 2005; Fischer et al., 2007). They are thought to occur in the deep $H_2O$ clouds that exist in these atmospheres and are estimated to be $10^3$ to $10^6$ times more energetic than on Earth (Yair et al., 1998; Gurnett et al., 2005).

**1.2 Transient Luminous Events (TLEs)**

Transient Luminous Events (TLEs) is the collective name given to a wide variety of optical emissions which occur in the upper atmosphere above active thunderstorms. These very brief (from few to tens of ms), colorful phenomena were discovered serendipitously in 1989 (Franz et al., 1990), and since then have been extensively studied from the ground, aircraft, balloons, the space shuttle, the International Space Station and from various orbiting satellites. There is now considerable theoretical and observational literature that covers the phenomenology, morphology and relationship to lightning parameters of TLE generation (see for example the monograph edited by Füllekrug, Rycroft and Mareev, 2006, the review by Neubert et al. (2008) or the cluster issue of J. Phys. D on streamers, sprites and lightning edited by Ebert and Sentman, 2008). Distinct classes and names - Elves, sprites, carrots, columns, angels, jets, trolls and pixies - were given to the various shapes of these optical emissions – all of which occur in mesospheric and stratospheric heights above active thunderstorms. Sprites are the most spectacular type of TLE, with shapes reminiscent of giant jellyfish, carrots or columns and they are usually associated with intense positive cloud-to-ground lightning (+CG) that posses a large charge-moment change (Cummer and Lyons, 2005). The suggested mechanism for sprite production is the emergence of a quasi-electrostatic (QE) field between the cloud top and the ionosphere after the cloud's positive charge center is removed from the thundercloud. This QE-field can be sustained for a time of the order of magnitude of the local relaxation time (which depends on the local conductivity), and it is sometimes strong enough to exceed the conventional breakdown field in mesospheric heights (Pasko et al., 1997). Thus sprites can be delayed by ~tens of milliseconds from the parent lightning (Sao-Sabbas et al., 2003) and they are usually initiated in the height range of 70-80 km, from which they propagate in visible tendrils downwards and upwards. High-speed imagery (Stenbaek-Nielsen et al., 2007; McHarg et al., 2007) showed that sprites start with downward moving streamer heads which have sizes of the order of hundred meters or less. Sprite brightness can reach several mega Rayleigh (Gerken et al., 2003), lasting up to several tens of milliseconds. The main emissions in sprites include the first positive ($N_2$1P) and second positive ($N_2$2P) band systems of $N_2$, the $N_2$ Lyman–Birge–Hopfield (LBH) band system and the first negative band system of $N_2^+$ ($N_2^+$1N) (Morrill et al., 2002). The molecular basis for various sprite emissions is reviewed in detail by Pasko (2007) and by Sentman et al. (2008a, b). Sprites seem to play an important role in mesosphere-troposphere coupling that bears on the global electrical circuit (Rycroft et al., 2002). Several observations suggest that certain types of jets that evolve into sprites may connect the top of thunderstorms directly to the ionosphere (Pasko et al., 2002). The potential role of sprites in altering the chemical properties of the mesosphere was recently reviewed by Gordillo-Vázquez (2008). There are few reports that sprites were produced by negative cloud-to-ground flashes (Barrington-Leigh et al., 2000) or by Intra-cloud (IC) discharges (Ohkubo et al., 2005), but these are probably rare exceptions. The inhomogeneity and transient variability of the terrestrial atmosphere at mesospheric heights is believed to play a crucial role in the initiation of TLEs - e.g. gravity waves, chemical reactions and meteor ablation products modify the local electrical properties of the mesosphere, making it more conducive to electrical breakdown processes. Sprites, haloes and elves seem to be ubiquitous around the planet, and were reported over most major centers of lightning activity on Earth. Space-based observations offer a global vantage point for studying the planetary occurrence rate of TLEs.

2. **Model Description**

Since lightning has been found in other planetary atmospheres, it seems reasonable to assume that some form of TLEs may also occur on those planets. Detailed calculations of the conventional breakdown parameters for various planetary atmospheres have been presented lately by Roussel-Dupré et al. (2008), for a range of external electric fields. Here we present initial calculations of the necessary lightning induced charge-moment changes and possible atmospheric heights for the occurrence of sprites on Venus, Mars, Titan and the gas giants Jupiter and Saturn. The calculations of the breakdown parameters and the critical electric field $E_k$ as a function of atmospheric pressure for the atmospheres of several types of solar system objects are based on Sentman (2004). In that work, detailed Boltzmann modeling of the electron distribution function as a function of the pressure-normalized electric field [E/p] was used to calculate $E_k$ values for various atmospheric compositions, representing the major planets. For simplifying the computation Sentman (2004) classified planetary atmospheres to 4 types based on their major constituents (e.g $N_2/O_2$, $CO_2/N_2$, $H_2/He$ and $N_2/CH_4$), corresponding to Earth, Venus/Mars, the giant planets and Titan/Triton, accordingly. The ionization (α) and attachment (η) coefficients and the electron mobility (μ) were calculated as a function of the reduced electric field over the range $10 \leq E/p \leq 100$ [V cm$^{-1}$ torr$^{-1}$] and T=300K, using the following analytical approximations:

$$\frac{\alpha}{P} = A_i \exp\left(-\frac{B_i}{E/P}\right) \qquad \frac{\eta}{P} = A_a \exp\left(-\frac{B_a}{E/P}\right)$$

where the coefficients $A_i$, $B_i$, $A_a$ and $B_a$ were taken from Sentman (2004), and P is the ambient pressure and E is the electric field strength. The field where the ionization rate just surpasses the attachment rate for given pressure and temperature, is called the conventional break-down field $E_k$; it should not be confused with the field required for streamer propagation after emergence which is lower. We calculated the values of $E_k$ over a wide range of pressures and temperatures in each planet's atmosphere, assuming that sprites occur below the base of the ionosphere and above the uppermost planetary cloud layer. As the determining physical quantity is the reduced field E/N, where N is the number density of molecules in the respective atmosphere, we used the ideal gas law to derive the E/N values from the E/p values at T=300 K .

In its simplest form the QE model assumes a thundercloud with a dipole charge structure where positive charge resides above the negative lower one. Since this structure was generated through slow charge separation, the cloud at this stage is electrically neutral and the electric field above it decays rapidly. After the positive charge is removed by a cloud-to-ground flash, the remaining negative charge produces an instantaneous large quasi-electrostatic monopole field. The conducting surface of the earth immediately screens this electric field; the surface-thundercloud system at the altitude of the mesosphere can then be approximated by a dipole field (see below). As an elementary calculation shows, this field at mesospheric heights and above is characterized by the charge moment change (CMC), i.e., the charge times its height above the surface (Wilson, 1929). In physical terms, the charge moment is the dipole moment of the cloud-earth-system after the lightning stroke; it is the leading order term in a multipole expansion (Jackson, 1962).

This field endures for a time equal or shorter to the local electrostatic relaxation time in each altitude given by $\tau_\sigma(z) = \varepsilon_0/\sigma(z)$, where $\sigma(z)$ is the height-dependant conductivity. For typical mesospheric values, at 70 km $\sigma(z)=10^{-7}$ S m$^{-1}$ and $\tau_\sigma(z)$ is equal to $10^{-4}$ s. The transient field accelerates ambient electrons and leads to excitation and ionization, and eventually to the formation of a sprite with its characteristic emissions. A thorough discussion of this approach can be found in Pasko et al. (1997) and Raizer et al. (1998). In order to compute the expected field above thunderclouds in other planets, we place a charge equivalent to the uppermost charge center expected in each cloud-type.

Then the electrostatic field for a charge Q, located at an altitude $Z_s$ above the conducting surface at altitude 0 is in three-dimensional space (x,y,z):

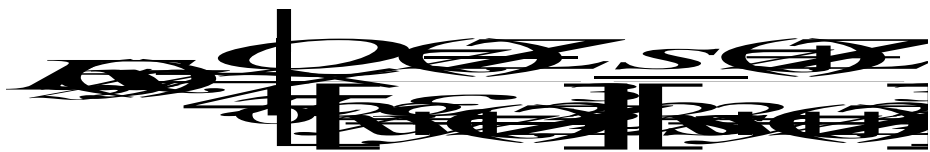

This approximation was given by Raizer (1998). The first term accounts for the cloud charge while the second accounts for the equipotential earth represented by the image charge trick (Jackson, 1962). On the axis of the cloud-earth-dipole (x=y=0), and neglecting the contribution of the ionosphere, the vertical component $E_z$ at altitudes where $z >> Z_s$ reduces to

$$E_z = \frac{Q}{2\pi\varepsilon_0} \left[ \frac{1}{(z-z_1)^2} - \frac{1}{(z-z_2)^2} \right]$$

Similar expressions were used by Pasko et al. 1997, 1998; Hiraki and Fukunishi, 2006; and many others for terrestrial sprites.

The model represents the charge in the thundercloud by point charges which is a valid approximation at heights much larger than the extension of the charged regions; it does not describe the detailed temporal evolution of the breakdown process and the evolution of sprites as the models of Pasko et al. (1998, 2001). Rather, we postulate a set of simple charge configurations in various types of clouds on other planets. This approach oversimplifies the complex reality of the charge distribution so as to account for the uncertainties in both charge locations and lightning discharge types in Venus, Titan and the giant planets, however, it takes the leading dipole field into account. The methodology is inspired by Wilson's (1925) reasoning for predicting that field-induced electrical breakdown can occur high in the terrestrial atmosphere (80 km): where $E_s$ equals to or surpasses the critical breakdown field $E_k$, the atmosphere can produce a discharge (e.g. sprites in present day terminology) if the right concentration of free electrons is present. The ubiquity of galactic cosmic ray- and solar EUV-induced ionization in the solar system ensures that this condition is likely to be met in all the planets we study. Updated discussions of the various breakdown processes possible under the influence of an external field are found in Cooray (2003) and Treumann et al. (2008). By definition $E_k$ is the homogenous electric field in which the ionization rate exceeds the attachment (de-ionization) rate, such that free electrons can grow into an avalanche and further into a streamer. This value is therefore an upper boundary for the initiation of a conventional breakdown process, but it is well known that if there are field inhomogeneities like in the neighborhood of particles (or laboratory electrodes), the field can exceed $E_k$ locally and start a discharge while it is below $E_k$ further away. Once created, a streamer discharge enhances the electric field at its tip (Raether, 1939) and therefore can propagate into regions where the field is well below $E_k$.

The vertical pressure and temperature profiles needed for calculation of the values of $E_k(z)$ were taken from the Planetary Atmospheres Node of the Planetary Data System (PDS) (http://atmos.nmsu.edu/) for Venus, Mars and Titan, and from Stanford's standard atmosphere computations (http://aero.stanford.edu/StdAtm.html) for the case of Earth's atmosphere. Although little is known about the locations and magnitudes of charge centers in the thunderclouds on other planets, thermodynamic models and abundance constraints allow us to postulate the possible altitudes and depths of clouds in various planets (Atreya, 1986). Early modeling work further constrains the location of electric charge centers in the different planets (Yair et al., 2008, and references therein).

## 3. RESULTS

*a. Earth*

In order to calibrate the model prior to applying it to alien conditions we re-calculated the predicted occurrence heights of sprites in the terrestrial atmosphere, in a manner similar to that presented by Pasko et al. (1997). The thunderstorm charge distribution assumes that a negative charge is located at 5 km and an equal and opposite charge resides at 10 km, typical altitudes for sprite-producing winter thunderstorms in Japan and the eastern Mediterranean (Ganot et al., 2007). This dipole description significantly simplifies the multilayer charge structure, typical of sprite–producing summer supercells (Stolzenburg et al., 2001; Lyons et al., 2008), but for the purpose of the present study it is sufficient. After the positive cloud-to-ground flash that removes the upper positive charge center (+CG) the negative screening charge center remains at approximately cloud top, leading to the short-lived QE-field in the space between cloud top and the ionosphere. The results are presented in Figure, 1 for different values of Q. For an initial configuration of two 50C charges placed at 5 and 10 km (a postulated charge-moment change of 10 km x 50 C=500 C·km by the flash), the conventional breakdown field is exceeded at ~84 km, where the conventional breakdown field is 10.26 V/m. For a charge of 100C (CMC=1000 C·km) this threshold is achieved at 79 km ($E_k$=26.6 V/m). These values are in good agreement with Pasko et al. (1997) who used a 1000 C·km charge moment change for evaluating sprite onset altitudes.

*b. Venus*

Venus is a terrestrial planet completely covered by a thick layer of sulfuric acid clouds in three layers spanning the 50-70 km altitude range above the surface (Markiewicz et al., 2007)). The ionospheric layer peaks at 140 km with an electron density ~4·10$^5$ cm$^{-3}$ (Pätzold et al., 2007). The huge surface atmospheric pressure (~9 MPa) renders the existence of cloud-to-ground discharges from the high-level sulfuric acid clouds ($Z_s$ ~ 50 to 70 km) practically impossible, due to the high value of the required field for electrical breakdown. Thus it is highly likely that lightning activity on Venus would be comprised entirely of intra-cloud (IC) discharges, either between clouds at the same layer or between two separated cloud decks. Based on the Cassini spacecraft data, Gurnett et al. (2001) stated that such IC discharges would be slow to build-up and would exhibit different characteristics compared to their terrestrial analogues. A slow discharge would meet the requirement of the QE-model (Pasko et al., 1997) and based on the above-cloud conductivity profile used by Pechony and Price (2004) the penetration of thundercloud induced electric fields into the upper atmosphere would last the order of 0.1 s before electrostatic relaxation. The potential occurrence altitudes of sprites for several values of charge, placed at a representative altitude of the cloud-deck at 50 km, are presented in Figure 2. The threshold for conventional breakdown is exceeded at 90 km (729 V/m) and 84 km (2862 V/m) for charges of 100 and 250 C, respectively. Thus, for an IC flash with a charge-moment change of 100 C ·5 km =500 C·km occurring between the two lower cloud layers (presumably separated by 5 km), a sprite can be triggered approximately at an altitude 90 km above ground, ~20 km above the tops of the upper most cloud layer. As can be expected, larger accumulation of charge would result in a sprite at lower altitudes and closer to the cloud tops, leaving just a narrow margin for attempted above-limb spacecraft observations.

The climatology of lightning activity in Venus is still largely unknown, and hence no latitudinal dependence of sprite occurrence can be prediceted. Since the clouds of Venus are assumed to posses a low mass content, it is hard to envision very large accumulations of charge due to the low efficiency of known charge separation mechanisms (Levin et al., 1983). Furthermore, based on the Venus Monitoring Camera on board Venus Express, the depth of sub-solar induced convection seems to be shallower than previously assumed (Markiewicz et al., 2007), and it does not penetrate into the middle and lower cloud decks. Still, this does not rule out the possibility that unknown types of charge generation processes do take place in the deep clouds of Venus that eventually lead to lightning and consequently to sprites.

c. *Mars*

Mars does not posses convective $CO_2$ or $H_2O$ clouds with sufficient vertical motions and particle concentrations to separate electrical charges and build-up sustained electric fields that will lead to conventional breakdown and lightning. Mars is frequented by dust storms, from local convectively unstable dust devils to regional and planetary scale events. Cantor et al. (1999) showed that dust storms are frequently spawned at the edges of the two polar cap, at the base of elevated regions in the northern Martian hemisphere, near the polar hood during northern fall and at mid-latitudes in both hemispheres. Charging in dust storms that occur in the lower atmosphere was simulated in laboratory experiments and modeled extensively, and it is expected that significant charge is generated within them by triboelectric charging (Sternovski et al. 2002; Kraus et al., 2003) and by attachment of galactic cosmic ray-induced ionization products (Michael et al., 2008). The numerical simulation by Melnik and Parrot (1998) suggested that this charging process is very efficient and generates local fields within the dust devil of ~20 kV/m which are sufficient to lead to breakdown (it should be pointed out that the coefficients used by Melnik and Parrot (1998) for $CO_2$ breakdown were based on Paschen parameters, which may not be applicable in large gap discharges such as expected inside dust storms with several km size). Farrell et al. (1999) maintain that larger storms, of the order of ~5 km, are unable to sustain large amounts of charge and will initiate corona and glow discharge. Our calculations show that even for the low case of Q=10C located at 10km above the surface, the conventional breakdown is surpassed within the dust cloud and not above it. Larger charges will have a similar effect, suggesting that sprite-like discharges on Mars will not occur above the charge center.

d. *Titan*

Titan is the only moon in the solar system that has a thick atmosphere, which is dominated by nitrogen and methane and has a surface pressure of 1.5 bars and a temperature of 94K.. A recent review by Lorenz (2008) describes the complex hydrological circle on Titan, with convective methane clouds producing occasional flash floods. Such clouds were intermittently observed over Titan's South Pole in 2004 and 2005 (Rannou *et al.*, 2006) and are assumed to contain sufficient mass-loading for particle interaction to be significant. The conductivity of Titan's atmosphere at altitudes above the troposphere was studied by Borcuki et al. (1987; 2006) and Whitten et al. (2007), and is often modeled as increasing with height to values $\sim 10^{-8}$ S m$^{-1}$ at 80 km implying a relaxation time of 1 ms for any lightning-induced QE field. We assumed a convective methane cloud possessing a monopole charge structure, with the negative charge center located 20 km above the ground, in accordance with the numerical simulations of cloud charging and lightning generation presented by Tokano et al. (2001). We used 3 different values of total charge: 50C, 100C and even the highly unrealistic amount of 250C. The results (Figure 3) show that electrical breakdown may be achieved within the cloud (indicating the possibility for lightning) but the resulting above-cloud QE electric field fails to surpass the conventional breakdown value even at an altitude of 200 km, meaning that no sprites can be expected in that atmosphere.

e. *Jupiter*

Jupiter is a gas giant with a hydrogen-helium atmosphere, and lacks a solid surface. The ionosphere is found approximately 1000 km above the 1-bar level (which is commonly used as a reference even though it lacks any special meteorological significance) with electron concentrations $\sim 10^5$ cm$^{-3}$, and a second peak at ~2000 km with $\sim 10^4$ cm$^{-3}$ electron concentration (Hinson et al., 1997). Inevitably, all lightning activity in Jupiter is intra-cloud, believed to originate within the deep convective water cloud layer (Williams et al., 1983). These mixed-phase clouds can be several tens of kilometers deep and can be very efficient in charge separation, able to rapidly generate strong electric fields that surpass the critical value and produce very energetic lightning (Yair et al., 1995). The inner boundary (in terms of the electrical capacitor concept of the planetary global circuit it is a perfect electric conductor) is determined by the conductivity profile and the corresponding skin depth. In the ELF range, where the lightning produced Schumann Resonances are commonly used to detect electrical activity (Yang et al., 2006), a conductivity of ~0.01 S/m provides a skin depth of a few km, which is much less than cavity thickness. On Jupiter, such conductivity is reached at ~0.96 Jupiter radius which corresponds to ~150 kbar, very deep below the H$_2$O clouds which are located between 5 and 2 bars (Simoes et al., 2008). In this case equation (2) can be simplified to that of a monopole in free-space. For the model calculations of the above-cloud electrostatic field for a Jovian thundercloud we placed a single charge center containing 100, 500 and 1000 C within the deep H$_2$O thundercloud which is supposed to reside between 5 and 3 bars (Yair et al., 1995). The results suggest that for a 1000C charge located 30 km below the 1-bar pressure level (where Gibbard et al. (1995) and Yair et al. (1998) showed the upper charge center in the H$_2$O cloud reside) a sprite can be ignited at an altitude ~280 km above the 1-bar level, approximately 100 km above the top visible NH$_3$ cloud layer (Figure 4). A lower charge accumulation in the thundercloud (100C) will be manifested in higher generation altitudes for Jovian sprites, still below the ionosphere. Since lightning in Jupiter are absent in equatorial regions and are mostly located at mid- and high latitudes, we would expect Jovian sprites to appear above the clouds in these regions, whenever a strong convective system evolves. No seasonal dependence for such cloud structures had been reported.

**4. Expected emissions based on laboratory experiments**

The early experimental work of Borucki and McKay (1987) was conducted for predicting the optical efficiencies of various atmospheres in order to assess their capability to allow lightning light to emanate from the deep atmosphere so that it can be detected by a sensor on an orbiting spacecraft. The results showed that the fraction of the energy in lightning discharge channels that is radiated in the visible spectrum is similar for Earth, Venus and Titan, but quite different for Jupiter. Additional laboratory data on the possible spectrum of lightning in other planets was obtained by Laser-Induced Plasma (LIP) experiments conducted by Borucki et al. (1996). These showed the main spectral features of high-energy discharges in various gas mixtures representing the atmospheres of Venus, Jupiter and Titan. However, these experiments mimicked the high-current high-temperature lightning channel, and are different from the low-current low-temperature nature of sprites.

Experiments in low-temperature non-equilibrium discharges were conducted by Williams et al. (2006) who produced what they termed "sprite in a bottle" using DC currents inside an air-filled glow discharge tube at pressures of 0.01 to 1 Torr at room temperature. The current densities in the tube were ~1 mA m$^{-2}$ and the resulting spectra were in reasonable agreement with those observed in nature, as reported by Hampton et al. (1996). The radiance of the N$_2$1P line was found to be roughly proportional to the applied current density, where a 10 mA m$^{-2}$ current density corresponds to a brightness of 1 MR, considered in the upper limit for sprite brightness (Figure 8 in that paper). Similar experiments were conducted by Goto et al. (2007) for different atmospheric pressures, ranging between 0.0075 and 7.0 Torr (~10$^{-2}$ mb, approximately equivalent to 80 km altitude, to 9.33 mb which is equivalent to 31 km altitude). They used a high voltage electrode suspended inside a Pyrex glass tube, with applied DC voltages of 50 and 100 kV. The obtained spectra between 550 and 800 nm showed a reasonable agreement with the results of Hampton et al. (1996). The spectral distributions were shown to be independent of the relative humidity of the air, and displayed clear pressure dependence, with many lines becoming weak or disappearing as the pressure increases. In a recent work, Ohba et al. (2008) repeated the experiment for CO$_2$ in order to simulate the Venusian atmosphere between 5 and 100 Torr (6 to 133 mb), with 2 different voltages, 100 and 700 kV. At the high voltage, CO$_2^+$ bands were present at 354.5, 383.8, 404.8, 426.8, 434.4, 465.9, 512.9 and 543.0 nm, as well as the Hα (656.2 nm) and the OI line in 777.6 nm. These emissions are in good agreement with the early experiments of Smyth (1931) that showed several emission bands of the Fox-Duffendack-Barker system of CO$_2^+$ between 290 and 500 nm, with prominent lines near 288, 405 and 427 nm. In Venus, other emission lines from CO and OH are also possible. This suggests that the dominant colors of Venusian sprites may be blue-green and red. From theoretical considerations, we can expect that for the Hydrogen-Helium atmospheres of Jupiter and Saturn strong sprite emissions from atomic Hydrogen in 656 nm and from He near 447, 587 and 706 nm (based on the NIST database atomic spectra). Molecular hydrogen emissions may also be present, such as the Lyman and Werner bands in the EUV. Sprites on Titan, should they exist, would display the same spectra of terrestrial ones, since molecular nitrogen is the main component in both atmospheres.

These glow discharge experiments are physically similar to sprites, as in both cases the emissions are due to the collision of field accelerated electrons with gas molecules at room temperature. Therefore similar spectral lines occur. However, the electric field in the positive column of a long glow discharge is much lower than the enhanced field at the head of a streamer; in particular, these fields are well below or well above $E_k$. Images with high temporal resolution show the same glowing streamer heads in experiments (Ebert et al., 2006) and in sprite observations (Stenbaek-Nielsen et al., 2007 and 2008). The emissions from these heads are characteristic for the high local fields and electron energies in the streamer head, and therefore the relative intensity of spectral emissions differs between glow and streamer discharges. Another difference is due to the fact that in a stationary DC glow discharge at low fields, excitations can occur in two or more steps and are therefore strongly pressure dependent, while two-step processes are unlikely in the rapidly propagating high field zone at a streamer head, and spectral emissions therefore depend much less on pressure or density. Therefore the glow discharge experiments give an indication on the expected emission lines, but relative line intensities will have to be deduced from time resolved streamer experiments, or from theory.

**Discussion**

Model calculations predict the appearance of sprites in the upper atmospheres of Venus and Jupiter, based on reasonable assumptions on the amount of charge that may be present in the hypothetical thunderclouds in their atmospheres. The resultant spectral emissions for sprites in the atmospheres of other planets can be searched for by orbiting spacecraft observing the night side of the planet toward the limb, as was practiced during the MEIDEX sprite campaign from the space shuttle (Yair et al., 2003) and as being operationally conducted by the ISUAL instrument on board the FORMOSAT-2 satellite (Kuo et al., 2005; Cummer et al., 2006). In these missions, sprites were recorded above the Earth's limb from ranges of 1800-3000 km. This is a consequence of the low absorption of sprite light by the atmosphere that enabled enough photons to reach the on-board detectors. During the MEIDEX, the measured brightness of sprites was in the range of 0.3–1.7 mega-Rayleigh (MR; one Rayleigh is equal to $7.96 \cdot 10^4$ photons $s^{-1}$ $m^{-2}$ $sr^{-1}$) in the 665±50 nm range and 1.44–1.7 MR in the 860±50 nm range (Yair et al., 2004), typical of emissions of the $N_2$1P group.

For Venus, the Japanese Planet-C mission (Nakamura et al., 2007; Takahashi et al., 2008) is designed to carry the Lightning and Airglow Camera (LAC) which will be used to observe lightning on Venus even if they are less bright by a factor of 100 compared to terrestrial flashes, when viewed from a 100km altitude. The LAC will have multispectral capabilities and will cover the oxygen lines in 777.4 nm (OI, related to lightning) and in 552.5 nm [$O_2$ Herzberg II], 557.7 nm [OI] and 630.0 [OI], which are airglow emissions. Since the field of view is 16º, it is highly likely that any sprite emission above the cloud tops will be detected by the instrument as it observes the limb. The fact that no optical emissions from Venus lightning had been thus far reliably detected means that they may be weak or completely hidden due to absorption by the high-level clouds. This poses an interesting challenge for the Planet-C mission, namely, to indirectly infer the existence of lightning in Venus based on their sprite fingerprints. For Mars, the possibility that corona or other types of discharges occur inside dust devils and in large-scale dust storms can be optically verified by sensitive cameras on board future landers. The possibility that such discharges can be remotely sensed by orbiting spacecraft is unclear, due to absorption and obscuration by the dust. However if sparks do emanate from the upper parts of the dust clouds upwards into the free atmosphere (upward flashes had been reported lately on Earth; Krehbiel et al., 2008) they can be sought for by night-time oblique observations of the dust column, in order to allow escaping photons to reach the sensor.

The duration and morphology of alien sprites will strongly depend on the nature of the parent lightning and on the properties of the atmosphere. Once the critical breakdown field had been surpassed locally, electron avalanches can evolve into streamers if the ambient (pre-breakdown) electron density is appropriate; the maximal local field depends on the temporal behavior of the lighting strokes and on spatially varying dielectric relaxation times, which in turn depend on the ambient conductivity. These conditions are discussed in detail by Pasko et al. (1998, 2000) for terrestrial sprites. We presently have little information on these parameters for other planets, yet some insights can be deduced from observations and laboratory experiments that use high-speed photography to study electrical sparks and streamers in various gas mixtures, based on similarity laws and the analogy drawn between sprites and gas-discharge sparks. Stenbaek-Nielsen et al. (2007) and McHarg et al. (2007) have shown that the tips of sprites are glowing balls that move rapidly ($10^7$ m/s.) and typically brighten as they travel up or down, with brightness exceeding 60 MR. The same structure was seen at the tips of streamers produced in laboratory experiments (Ebert et al., 2006). Recently, Briels et al. (2006) reported on the properties of streamers with diameters varying gradually between 0.2 and 2.5 mm, as recorded by an iCCD camera with nanosecond resolution. In another work Briels et al. (2008) showed that similarity laws apply for the morphology of streamers at varying gas densities, and Nijdam et al. (2008) used the same system to study the 3D structure and the branching of streamers. In an appropriate vacuum chamber, such experiments can be replicated in the appropriate planetary gas mixtures, in order to elucidate the nature of sprites occurring in these atmospheres. Present measurements in pure nitrogen and pure argon already indicate that streamer properties (inception, width, length, branching) largely depend on gas composition. This will be a subject for future research.

**Figure Captions**

**Figure 1**: The electrostatic field above typical terrestrial wintertime sprite producing thundercloud. Monopole structure with 50, 100 and 250 C located at 10 km, respectively. Where the conventional breakdown field (brown) crosses the E-field, a sprite can potentially occur.

**Figure 2:** The electrostatic field above a Venusian charge configuration placing opposite and equal charges at the lower (40km) and middle (50km) cloud decks. The critical breakdown field is surpassed at 92 km (for 500C) and 100 km (for 100C).

**Figure 3:** The electrostatic field above a Titan thundercloud possessing a negative charge located at 25 km. Even at high values of total charge the breakdown field is not exceeded.

**Figure 4:** The electrostatic field above a Jovian thundercloud possessing a negative charge located 30 km below the 1-bar pressure level. The field is exceeded for a realistic charge of 1000 C at an altitude of 280 km above the 1-bar level.

# Figures

Figure 1

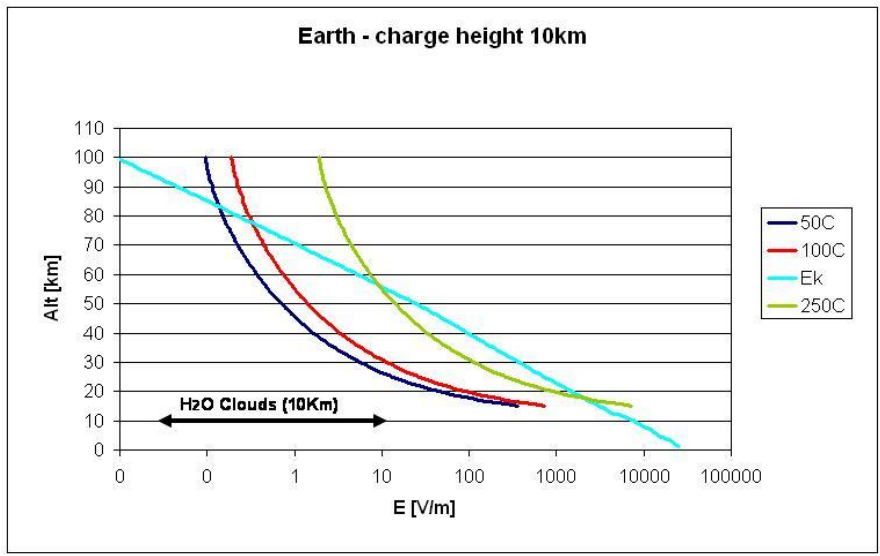

Figure 2

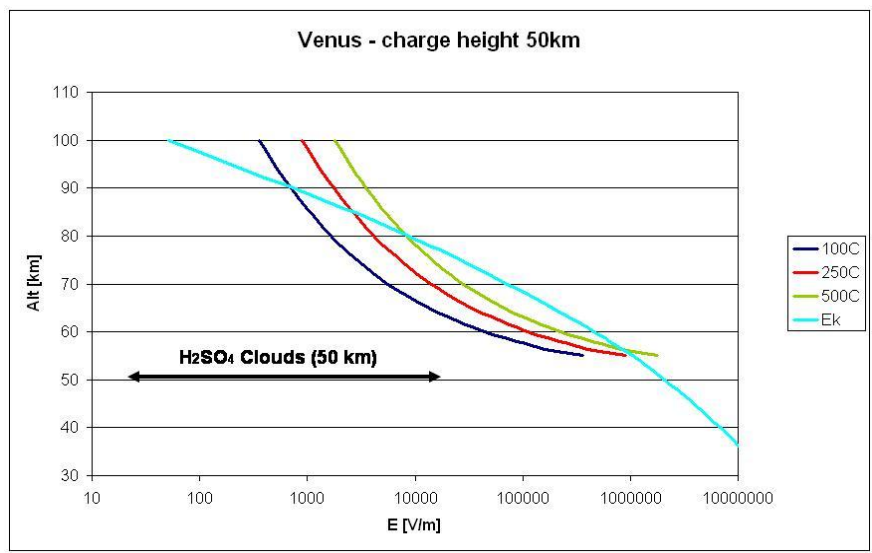

Figure 3

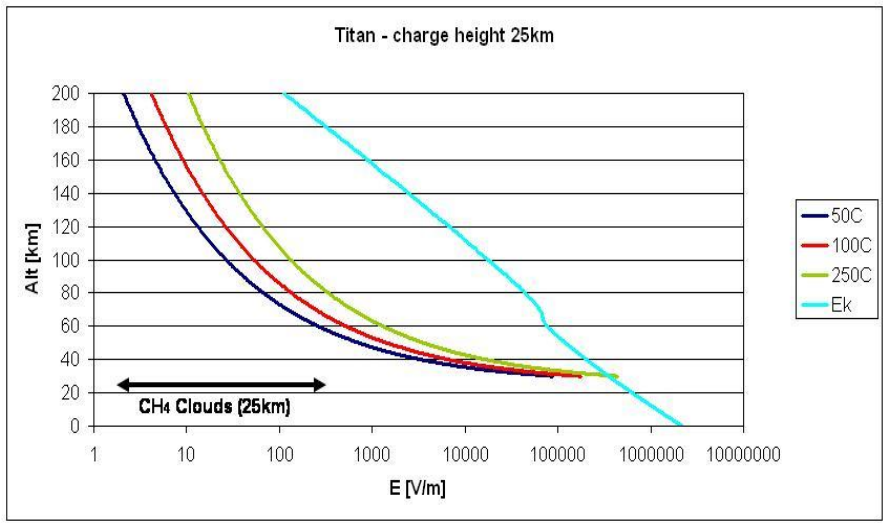

Figure 4

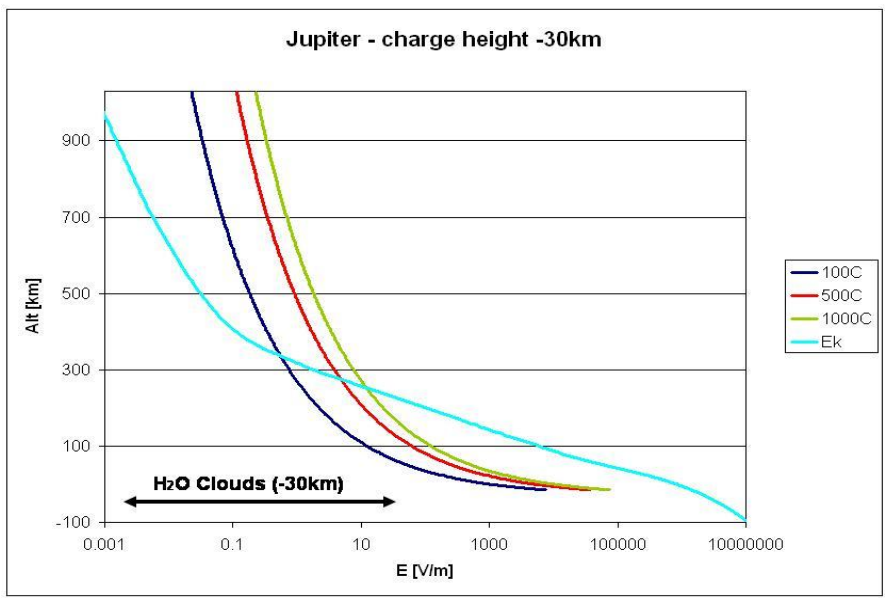